\documentclass{ws-ijmpa}
\usepackage[super,compress]{cite}
\usepackage{graphicx}
\usepackage{color}
  \newcommand{\nn}{\nonumber}
\begin{document}
\markboth{T.Kitadono and T. Inagaki} {Elliptically oscillating solutions in Abelian-Higgs model} 

\catchline{}{}{}{}{}
%
%%%%%%%%%%%%%%%%%%%%%%%%%%%%%%%%%%%%%%%%%%%%%%%%%%%%%%%%%%%%%%%%%%%%

%---
\title{Elliptically oscillating solutions in Abelian-Higgs model and electromagnetic property}

%---
\author{Yoshio Kitadono\footnote{Corresponding author}
}

\address{Liberal Education Center, National Chin-Yi University of Technology, \\
No.57,  Sec.2, Zhongshang Rd.,Taiping Dist.,\\
 Taichung 41170, Taiwan, Republic of China (R.O.C.)\\
kitadono@ncut.edu.tw}

%---
\author{Tomohiro Inagaki}

\address{Information Media Center and Core of Research for the Energetic Universe, Hiroshima University,\\
No.1-3-2, Kagamiyama,Higashi-Hiroshima,
Hiroshima 739-8521, Japan 
\\
inagaki@hiroshima-u.ac.jp}

\address{International Laboratory of Theoretical Cosmology, Tomsk State University of Control Systems and Radioelectronics (TUSUR), 634050 Tomsk, Russia}

\maketitle

\begin{history}
\received{Day Month Year}
\revised{Day Month Year}
\end{history}

%--- Abstract ---
\begin{abstract}
The elliptically oscillating solutions in the Abelian Higgs-model are presented and the classical massive-dispersion-relation through the non-linear dynamics is discussed. The generated massive-dispersion-relation including a field value of the scalar field is derived as the consequence of the equation of motions. We discuss the property of the new solutions and its Hamiltonian density. In addition, we calculate the electromagnetic property of the system, in particular, we derive the relation between the field value and the electric field and the electric current-density.   

%---- Keywords ---
\end{abstract}

%--- PACS ---
\ccode{PACS numbers:~05.45.-a;14.80.Cp;15.15.Ex;11.10.Lm;03.50.-z}
% 05.45.-a: Nonlinear dynamics and chaos
% 14.80.Cp: Non-standard-model Higgs bosons
% 15.15.Ex: Spontaenous breaking of gauge theory
% 11.10.Lm: Nonlinear or nonlocal theories and models
% 03.50.-z: Classical field theories

%\tableofcontents

%--- Sec.1 Introduction
\section{Introduction \label{Sec1}}
The mechanism of giving masses to elementary particles is known as Higgs mechanism \cite{Englert1964,Higgs1964,Guralnik1964}. In this mechanism, the masses of elementary particles are given as the product of the vacuum expectation value (VEV) and the couplings between Higgs and each particle. The key of this mechanism is that VEV is determined by the minimum of the Higgs potential. The phenomenological consequences based on the standard model with this mechanism have been tested by many experiments, for example, the Large Hadron Collider and no serious discrepancy between the theory and data has been observed \cite{Higgs.ATLAS, Higgs.CMS}.

In quantum field theories, there is an interesting example giving massive-dispersion-relation through non-linear dynamics, namely, the  massive solutions for non-linear equation of motion (EOM). In 1970s, this property of the solution was already known in the literature \cite{Treat1971}, in the context of investigating a non-zero massive-dispersion-relation for the solution of EOM in $SU(2)$ pure non-Abelian Yang-Mills theory. The solution of the EOM in this system shows that the massive-dispersion-relation is generated as the consequence of the squared momentum in the solution of the EOM.

Coleman suggested an existence of non-linear-plane-wave in non-Abelian gauge theory in the end of 1970s \cite{Coleman1977}. It triggered the intensive study of non-linear-plane-wave solutions in non-Abelian Yang-Mills gauge theories. Corrigan and Fairlie pointed out so-called Corrigan-Fairlie-'t Hooft-Wilczek ansatz \cite{Corrigan1977} assuming a special form of the solution for Lorentz- and gauge-indexes of the gauge field in $SU(2)$ pure-Yang-Mills theory. This assumption leads the EOM of the gauge field to a simplified EOM of a scalar theory with $\phi^4$ potential. Oh and Teh extended the assumption of the gauge field in the pure $SU(2)$ gauge-theory to more general-forms \cite{Oh1979,Oh1985}.  The solutions are described by the Jacobi's elliptic-functions, in particular, $cn$-type solution often appears as the solution of EOM in $SU(2)$-pure-Yang-Mills theory \cite{Baseyan1979,Matinyan1981}. In addition, general families of the elliptic solutions were discussed \cite{Cervero1977, Actor1979} and the energy transportation in $SU(2)$ gauge-theory coupling to $SU(2)$-triplet-scalar-field was discussed~\cite{Brihaye1983}. One of common property in these solutions is that the solution is described by one of Jacobi's elliptic-functions.

Relatively recently,  another type of the solution expressed by Weierstrass elliptic-function with an angular momentum in pure $SU(3)$ gauge-theory was discovered \cite{Tsapalis2016}. Phenomenologically, Achilleos et al. discussed a relation between stability of non-linear-plane-wave and Higgs mass in $SU(2)$ gauge field which couples to Higgs field \cite{Achilleos2011}. In addition, Frasca applied the Jacobi's $sn$-type-elliptic-function to the infra-red behavior of the gluon propagator \cite{Frasca2008} and also discussed a correspondence between massless $\phi^4$ theory to Yang-Mills gauge theory \cite{Frasca2009}. 

We applied the Jacobi's $dn$-type-elliptic-function to described the classical behavior of the Higgs field \cite{Kitadono2016}.  In the work, the authors did not take into account couplings of Higgs field to other fields.  However, if we consider a coupling between the Higgs field to other gauge-fields, how the solutions are affected by the gauge field ? and what kind of solution appears in the non-linear EOM ? To answer these questions by finding new solutions will be theoretically attractive and it will be helpful to expand the knowledge about non-linear dynamics of Higgs potential and gauge theories.  In addition, recently, new solutions in non-linear EOM of Abelian-Higgs model are discussed \cite{Mohammedi2021}, however the solution has the non-standard polarization. Thus we present the new solutions with the standard polarization in the Abelian-Higgs model. 

 In this paper,  we find out the new solutions with the standard polarization in non-linear EOMs of Abelian-Higgs model. One of the reasons why we consider Abelian-Higgs model is because that this model is the simplest extension of $\varphi^4$ theory with the gauge interaction. Another advantage to analyze this system is that finding observables is quite easy since the electromagnetic fields are obviously gauge-invariant quantities. 

In Sec.~\ref{Sec2}, we shortly review the Abelian-Higgs model and discuss the EOMs with the gauge fixing. We focus on some electromagnetic properties in this theory in Sec.~\ref{Sec3}. We derive the analytic formulas for the electromagnetic field and the densities of electric-charge/electric-current. We summarize the work in Sec.~\ref{Sec4} by emphasizing a way connecting the generated massive-dispersion-relation through the non-linear dynamics to observables.

%--- Sec.2 The Model and Solutions
\section{Abelian-Higgs model and elliptic solutions \label{Sec2}}
\subsection{Equation of motions}
The Abelian-Higgs model is described by the following Lagrangian \cite{Higgs1964} (alternatively see textbooks \cite{Peskin,Itzykson}):
\begin{eqnarray}
\mathcal{L} &=& (D_{\mu} \phi)^{*}(D^{\mu}\phi) - V(\phi) - \frac{1}{4}F_{\mu\nu}F^{\mu\nu}, \label{lagrangian-1}\\
%---
 V(\phi) &=& - \mu^2(\phi^{*}\phi) + \lambda(\phi^{*}\phi)^2,
\hspace{1cm}
F^{\mu\nu} = \partial^{\mu}A^{\nu} - \partial^{\nu}A^{\mu}, 
%---
\end{eqnarray}
where $D_{\mu}=\partial_{\mu}+ieA_{\mu}$ is the covariant derivative, $e$ is the electric charge, $A_{\mu}$ is the Abelian gauge field, $\mu$ is the mass parameter, $\lambda$ is the self coupling, and $\phi$ is the Abelian-complex-scalar-field. As is well known, the Lagrangian in Eq.~(\ref{lagrangian-1}) is invariant under the gauge transformations:
\begin{eqnarray}
  \phi(x) &\to& \phi^{\prime}(x) = e^{-ie\theta(x)}\phi(x), \label{gauge.trans.phi}\\
 A_{\mu}(x) &\to& A^{\prime}_{\mu}(x) = A_{\mu}(x) + \partial_{\mu}\theta(x). \label{gauge.trans.A}
\end{eqnarray}
The scalar field $\phi$ can be parametrized by two real-fields, $\varphi$ and $\chi$,
\begin{eqnarray}
 \phi(x) = \frac{1}{\sqrt{2}}\varphi(x)e^{i\chi(x)},
\end{eqnarray}
then the Lagrangian in Eq.~(\ref{lagrangian-1}) reduces to
\begin{eqnarray}
 \mathcal{L} &=&
   \frac{1}{2}(\partial \varphi)^2 
 + \frac{1}{2} \varphi^2(\partial \chi)^2 
 + eA^{\mu}(\partial_{\mu}\chi)\varphi^2
 + \frac{1}{2} e^2 A^2 \varphi^2
 - V(\varphi)
 - \frac{1}{4} F_{\alpha\beta}F^{\alpha\beta}, \label{lagrangian-2}
\end{eqnarray}
with the potential term, $V(\varphi) =
 - \mu^2\varphi^2/2 + \lambda\varphi^4/4$. The gauge transformation for $\phi$ can be rewritten by those for $\varphi$ and $\chi$:
\begin{eqnarray}
 \varphi(x) &\to& \varphi^{\prime}(x) = \varphi(x), \label{gauge.trans.varphi}\\
%---
 \chi(x) &\to& \chi^{\prime}(x) = \chi(x) - e\theta(x), \label{gauge.trans.chi}
\end{eqnarray}
and then the Lagrangian in Eq.~(\ref{lagrangian-2}) is invariant under the above gauge transformations.

The EOMs for this Lagrangian can be easily derived by the standard manner and those reduce to the following forms:
\begin{eqnarray}
 0 &=& \partial^2 \varphi - \varphi \left(\partial^{\mu}\chi + e A^{\mu}\right)^2 -\mu^2 \varphi^2 + \lambda \varphi^3, \label{eom.varphi.1}\\
%---
0 &=& \partial_{\mu} 
\left[ \varphi^2\left(\partial^{\mu}\chi + e A^{\mu}\right) \right], \label{eom.chi.1} \\
%---
0 &=& \partial_{\nu}F^{\mu\nu} - e\varphi^2
\left(\partial^{\mu}\chi + e A^{\mu}\right), \label{eom.A.1}
\end{eqnarray}
where $(\partial^{\mu}\chi + e A^{\mu} )$ in the EOMs, is the invariant set under the gauge transformations in Eqs.~(\ref{gauge.trans.varphi}),(\ref{gauge.trans.chi}). For simplicity, we take the special gauge, $\chi=0$ and $\varphi\neq 0$, in this paper. Note that we should take into account Eq.~(\ref{eom.chi.1}) even if we take $\chi=0$ because it gives the non-trivial constraint.

Although it is difficult to obtain general solutions of the EOMs, one of the important solutions is an oscillating solution in the potential, because it describes a stable behavior of the system. As we mentioned earlier work in the introduction, Jacobi's elliptic functions have been studied in EOMs of gauge theories. Hence we consider the elliptic solutions as special solutions with the standard polarization vector in Eqs.~(\ref{eom.varphi.1})-(\ref{eom.A.1}), namely, our polarization $\epsilon_{\mu}$ and momentum of the gauge field satisfy the identities, $\epsilon_{\mu}\epsilon^{\mu}=-1$ and $p_{\mu}\epsilon^{\mu}=0$ (note that the author in Ref.~\citen{Mohammedi2021} discussed the solutions with the special polarization satisfying  $\epsilon_{\mu}\epsilon^{\mu}=0$). 

First, we assume that the solutions are described by the Jacobi's elliptic-functions, which are finite and regular, and it has a massive dispersion-relation with respect to $p^{\mu}$. There are two solutions which have these properties and these are described by $cn(u,k)$ and $sd(u,k)= sn(u,k)/dn(u,k)$ \cite{Oh1979,Oh1985}. One of the solutions expressed by $cn(u,k)$ can be obtained in the following way. We assume the following function form for the gauge field:
\begin{eqnarray}
 A_{\mu}(x) = a \epsilon_{\mu} cn\left(u,k\right), \hspace{1cm} u=p\cdot x + \theta. \label{eq.A.cn.solution}
\end{eqnarray}
where $a$ is an overall constant which will be fixed by the EOMs later, and $\theta$ is a constant which cannot be fixed in this method (note that this $\theta$ in Eq.~(\ref{eq.A.cn.solution}) is different from the $\theta$ in the gauge transformation in Eq.~(\ref{gauge.trans.chi})). Substituting Eq.~(\ref{eq.A.cn.solution}) to
Eq.~(\ref{eom.A.1}), then we obtain,
\begin{eqnarray}
 \varphi =\sqrt{ \frac{p^2}{e^2} 
 \left(1-2k^2+2k^2cn^2(u,k)\right)}.
\end{eqnarray}

Next, we rewrite the EOM for $\varphi$ in Eq.~(\ref{eom.varphi.1}) into the EOM for $\varphi^2$ to avoid the square root in the above relation, then the EOM reduces to
\begin{eqnarray}
  \partial^2\varphi^2
 - 2(\partial_{\alpha} \varphi)(\partial^{\alpha}\varphi)
 - 2\left(\mu^2 + e^2A_{\alpha}A^{\alpha}\right)\varphi^2
 + 2\lambda \varphi^4=0. \label{eom.varphi.2}
\end{eqnarray}
Taking into account the differential equation for $cn(u,k)$ \cite{Gradshteyn,Abramowitz},
\begin{eqnarray}
 \frac{d}{du}cn(u,k) = -sn(u,k)dn(u,k),\label{eq.cn.diff.eq}
\end{eqnarray}
and the identities of the elliptic functions:
\begin{eqnarray}
 sn^2(u,k)+cn^2(u,k)=1, \hspace{1cm}
 dn^2(u,k)+k^2sn^2(u,k)=1, \label{eq.cn.identity}
\end{eqnarray}
then Eq.~(\ref{eom.varphi.2}) reduces to:
\begin{eqnarray}
 0 &=&
{}~~2\left[\tilde{k}^2+1 - \frac{e^2a^2}{p^2} - \frac{\lambda}{e^2}(\tilde{k}^2+1) \right]cn^6(u,k) \nn\\
&{}& + 
\left[-5\tilde{k}^2 + 2\frac{\mu^2}{p^2}
+ 4 \frac{e^2a^2}{p^2}\frac{\tilde{k}^2}{\tilde{k}^2+1}
+ 6 \frac{\lambda}{e^2} \tilde{k}^2
\right]cn^4(u,k) \nn\\
&{}& 
+\left[ 4\tilde{k}^2 - 4\frac{\mu^2}{p^2}
- 2 \frac{e^2a^2}{p^2}\frac{\tilde{k}^2}{\tilde{k}^2+1}
- 6 \frac{\lambda}{e^2} \tilde{k}^2
\right]cn^2(u,k) \nn\\
&{}& 
+ \left[ 1-\tilde{k}^2 + 2\frac{\mu^2}{p^2} \frac{\tilde{k}^2}{\tilde{k}^2+1}
+ 2 \frac{\lambda}{e^2} \frac{\tilde{k}^4}{\tilde{k}^2+1}
\right], \label{eq.4conditions}
\end{eqnarray}
with $\tilde{k}^2=2k^2-1$. To determine the unknown parameters, $k$, $p^2$, and $a$, we solve the above EOM by setting the coefficients of each power of $cn$-type function to be zero, then we obtain the following results:
\begin{eqnarray}
 k &=& \frac{1}{\sqrt{2}},  \hspace{1cm}
 a  = \sqrt{\frac{p^2}{e^2} \left(1-\frac{\lambda}{e^2}\right) }, \hspace{1cm} \mu = 0.
\end{eqnarray}  

Note that EOMs require the constraint on the parameters in the Lagrangian, $\mu$, because the number of equations (four) are larger than the number of unknown parameters (three; $k$, $a$, $p^2$). Hence, this potential reduces to the potential without the quadratic term. In addition, we obtain the additional constraint on the parameter space for $\lambda/e^2$ in order to keep the positivity in the square root of $a$, namely, we require the condition, $\lambda/e^2<1$. For simplicity, we rewrite $\varphi$ as $\varphi=\varphi_0cn(u,k)$. Then the dispersion relation for $p^{\mu}$ is converted into the form, $p^2=e^2\varphi^2_0$. The free parameter $\varphi_0$ can be regarded as an initial condition of $\varphi$ in solving this non-linear-differential-equation. 

Finally, the solutions are summarized into the following forms:
\begin{eqnarray}
 \varphi(x) &=& \varphi_0 cn\left(u,\frac{1}{\sqrt{2}}\right), \label{eq.sol.varphi.cn}\\
 A_{\mu}(x) &=& \varphi_0\sqrt{1-\frac{\lambda}{e^2}} \epsilon_{\mu} cn\left(u,\frac{1}{\sqrt{2}}\right), \label{eq.sol.A.cn}\hspace{1cm}\mbox{for}~~\left(\frac{\lambda}{e^2}<1\right),
\end{eqnarray}
with the massive-dispersion-relation, $p^2=e^2\varphi^2_0 \equiv m^2$. Depending on a field value $\varphi_0$, these solutions have the massive-dispersion-relation even if the quadratic term in the Lagrangian, $\mu$, is zero. Note that these solutions satisfy Eq.~(\ref{eom.chi.1}). This solution described by the Jacobi's elliptic function, $cn(u,1/\sqrt{2})$, which often appears in the quantum field theory \cite{Tsapalis2016}. For readers who are not familiar with this solution, see Sec.III in Ref.~\citen{Tsapalis2016} and references therein, since they shortly explain a reason why the EOM for pure $SU(N)$ gauge-theory shows this solution. (Note that our notation of the elliptic function, $cn(u,1/\sqrt{2})$, equals to their $cn(u,1/2)$ in Ref.~\citen{Tsapalis2016}.)

One can find out another type of solution described by the Jacobi's  $sn$-type function which appears in old literature \cite{Treat1971}. However, this solution reduces to the Jacobi's $sd$-type function. Although one can carry out the same procedure which we have showed in obtaining $cn$-type solutions. we simply assume the following ansatz for the solutions:
\begin{eqnarray}
  \varphi(x) &=& \varphi_0 sn\left(u,k\right), \hspace{1cm}
 A_{\mu}(x) = a\epsilon_{\mu} sn\left(u,k\right),
\end{eqnarray}
where we use the same notation for $u$ as used in Eq.~(\ref{eq.sol.varphi.cn}). Substituting these functions into the EOMs in Eqs.~(\ref{eom.varphi.1}) and (\ref{eom.A.1}), we obtain $k=i$ for the modulus of $sn$-type solutions. Actually one can convert this $sn(u,i)$ function into $sd$-type solution by using the formula for the Jacobi's elliptic-functions, namely, $sn(u,ik)=1/\sqrt{1+k^2}sd(\sqrt{1+k^2}u,k/\sqrt{1+k^2})$ with $sd(u,k)=sn(u,k)/dn(u,k)$ \cite{Gradshteyn, Abramowitz}. 
Then, we obtain the following solutions:
\begin{eqnarray}
 \varphi(x) &=& \frac{\varphi_0}{\sqrt{2}} 
 \frac{sn\left(u,\frac{1}{\sqrt{2}}\right)}
      {dn\left(u,\frac{1}{\sqrt{2}}\right)}, \label{eq.sol.varphi.sd}\\
 A_{\mu}(x) &=& \frac{\varphi_0}{\sqrt{2}} 
 \sqrt{1-\frac{\lambda}{e^2}} \epsilon_{\mu}
 \frac{sn\left(u,\frac{1}{\sqrt{2}}\right)}
      {dn\left(u,\frac{1}{\sqrt{2}}\right)}
 , \label{eq.sol.A.sd}\hspace{1cm}\mbox{for}~~\left(\frac{\lambda}{e^2}<1\right),
\end{eqnarray}
with the massive-dispersion-relation $p^2=e^2\varphi^2_0=m^2$. (Note we absorbed the numerical factor $\sqrt{2}$ in the front of $u$ and redefined the momentum as $p^{\mu}$ so that it satisfies the above dispersion relation). Here the popularization $\epsilon^{\mu}$ is either among $(0,1,0,0)$, $(0,0,1,0)$, and $(0,0,0,1)$, or a linear combination of them with a proper normalization constant. These solutions have been discovered and studied in literature to investigate solutions of EOM for $SU(2)$ non-Abelian Yang-Mills theory, see Refs. \citen{Oh1979,Oh1985}. We confirmed that these functions appear in the EOMs for the coupled system, that is, Abelian-Higgs model.

%---
\subsection{Hamiltonian density}
Next, we show the conservation of energy in this system for both $cn$- and $sd$-type solutions. It is noteworthy that the elliptic solutions do not vanish at the spatial infinity and hence one cannot use the standard technique based on the partial derivative to show the conservation of energy in the theory \cite{Peskin,Itzykson}. Thus, we directly show the constant nature of the Hamiltonian density.

The Hamiltonian density $\mathcal{H}$ in the system for $cn$-type solutions is defined by
\begin{eqnarray}
 \mathcal{H} = \pi_{\varphi}\dot{\varphi} + \pi^{\mu}_{A}\dot{A}_{\mu} - \mathcal{L},
\end{eqnarray}
where $\pi_{\varphi}=\partial\mathcal{L}/(\partial\dot{\varphi})$ and $\pi^{\mu}_{A}=\partial\mathcal{L}/(\partial\dot{A}_{\mu})$ are the standard canonical-momenta for each field. We can rewrite the above Hamiltonian-density into the following form,
\begin{eqnarray}
 \mathcal{H} = \mathcal{H}_{\varphi} + \mathcal{H}_{A} + \mathcal{H}_{\rm int}, 
\end{eqnarray}
where each of them is defined in:
\begin{eqnarray}
 \mathcal{H}_{\varphi} &\equiv& 
 \frac{1}{2}\pi_{\varphi}^2 + \frac{1}{2}\left(\partial_{i} \varphi\right)^2 + V(\varphi), \\\
 %---
 \mathcal{H}_{A} &\equiv& 
 \frac{1}{2}\pi_{A}^2 
 - \pi^{i}_{A}\partial^{i}A^0
 + \frac{1}{4}F^{ij}F_{ij},\\
 %---
 \mathcal{H}_{\rm int} &\equiv& - \mathcal{L}_{\rm int} = - \frac{1}{2}e^2\varphi^2A_{\mu}A^{\mu}.
\end{eqnarray}
To evaluate the Hamiltonian density, we take the rest frame of the momentum, namely, $p^{\mu}=(m,\vec{0})$ with $m=e\varphi_0$. 
We substitute the elliptic solutions into the above Hamiltonian-densities and then these three-contributions reduce to:
\begin{eqnarray}
\mathcal{H}_{\varphi}
&=& \frac{e^2}{2}\varphi^4_0 sn^2\left(u,\frac{1}{\sqrt{2}}\right)
dn^2\left(u,\frac{1}{\sqrt{2}}\right)
+ \frac{\lambda}{4}\varphi^4_0 cn^4\left(u,\frac{1}{\sqrt{2}}\right),\\
%---
\mathcal{H}_{A}
&=& \frac{e^2}{2}\varphi^4_0\left(1-\frac{\lambda}{e^2}\right)sn^2\left(u,\frac{1}{\sqrt{2}}\right)dn^2\left(u,\frac{1}{\sqrt{2}}\right),\\
%---
\mathcal{H}_{\rm int}
&=& \frac{e^2}{2}\varphi^4_0\left(1-\frac{\lambda}{e^2}\right)cn^4\left(u,\frac{1}{\sqrt{2}}\right).
\end{eqnarray}
One can easily show that the total Hamiltonian-density reduces to the following constant,
\begin{eqnarray}
\mathcal{H} = \frac{e^2}{2}\varphi^4_0\left(1-\frac{\lambda}{2e^2}\right), \label{eq.total.energy.density.cn}
\end{eqnarray}
where we used the identities for the elliptic functions in Eq.~(\ref{eq.cn.identity}).
The total energy $E$ is given as the spatial integral of the above Hamiltonian density, i.e., $E=\int \mathcal{H} d^3x = V\mathcal{H}$, with a finite spatial-volume $V$. Hence the conservation of the energy is obvious, i.e., $dE/dt=0$, as expected. 

The Hamiltonian density for $sd$-type solutions can be obtained by the exactly same procedure and each contribution of the Hamiltonian density reduces to:
\begin{eqnarray}
\mathcal{H}_{\varphi}
&=& \frac{e^2}{4}\varphi^4_0 
\frac{cn^2\left(u,\frac{1}{\sqrt{2}}\right)}
     {dn^4\left(u,\frac{1}{\sqrt{2}}\right)}
+ \frac{\lambda}{16}\varphi^4_0 
\frac{sn^4\left(u,\frac{1}{\sqrt{2}}\right)}
     {dn^4\left(u,\frac{1}{\sqrt{2}}\right)},\\
%---
\mathcal{H}_{A}
&=& \frac{e^2}{4}\varphi^4_0
\left(1-\frac{\lambda}{e^2}\right)
\frac{sn^2\left(u,\frac{1}{\sqrt{2}}\right)}
     {dn^4\left(u,\frac{1}{\sqrt{2}}\right)},\\
%---
\mathcal{H}_{\rm int}
&=& \frac{e^2}{8}\varphi^4_0\left(1-\frac{\lambda}{e^2}\right)
\frac{sn^4\left(u,\frac{1}{\sqrt{2}}\right)}
     {dn^4\left(u,\frac{1}{\sqrt{2}}\right)},
\end{eqnarray}
and the total Hamiltonian-density for $sd$-type solution coincide with Eq.~(\ref{eq.total.energy.density.cn}). 

\begin{figure}[htb]
%% CN solution
\hspace{-1.0cm}
\begin{tabular}{cc}
 \begin{minipage}[b]{0.45\linewidth}
  \centering
    \includegraphics[scale=0.35]{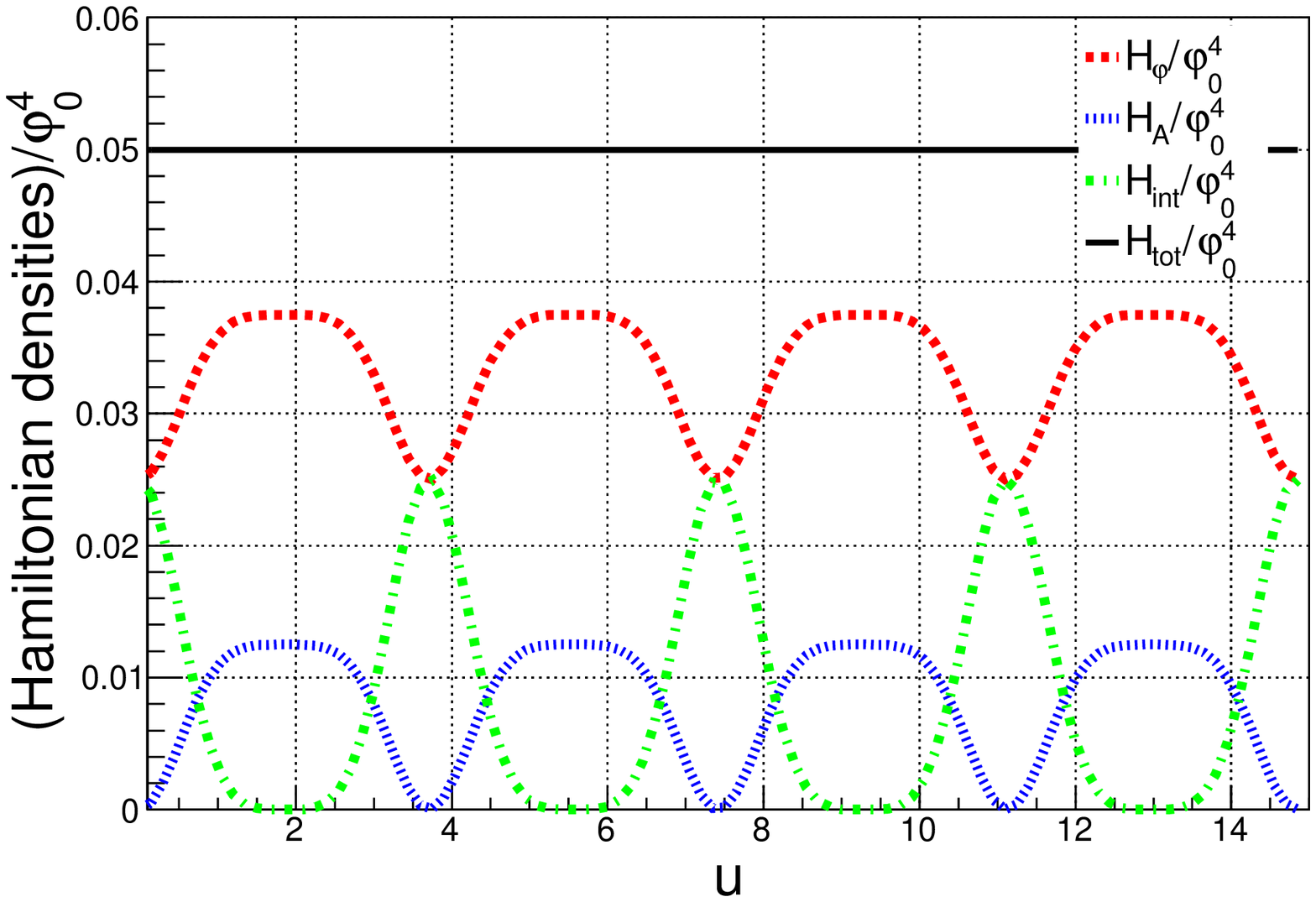}\\
  \centering
\hspace{1.5cm}
     (a) \\
 \end{minipage}
%% SN solucion  
 \hspace{1.cm}
 \begin{minipage}[b]{0.45\linewidth}
   \centering
     \includegraphics[scale=0.35]{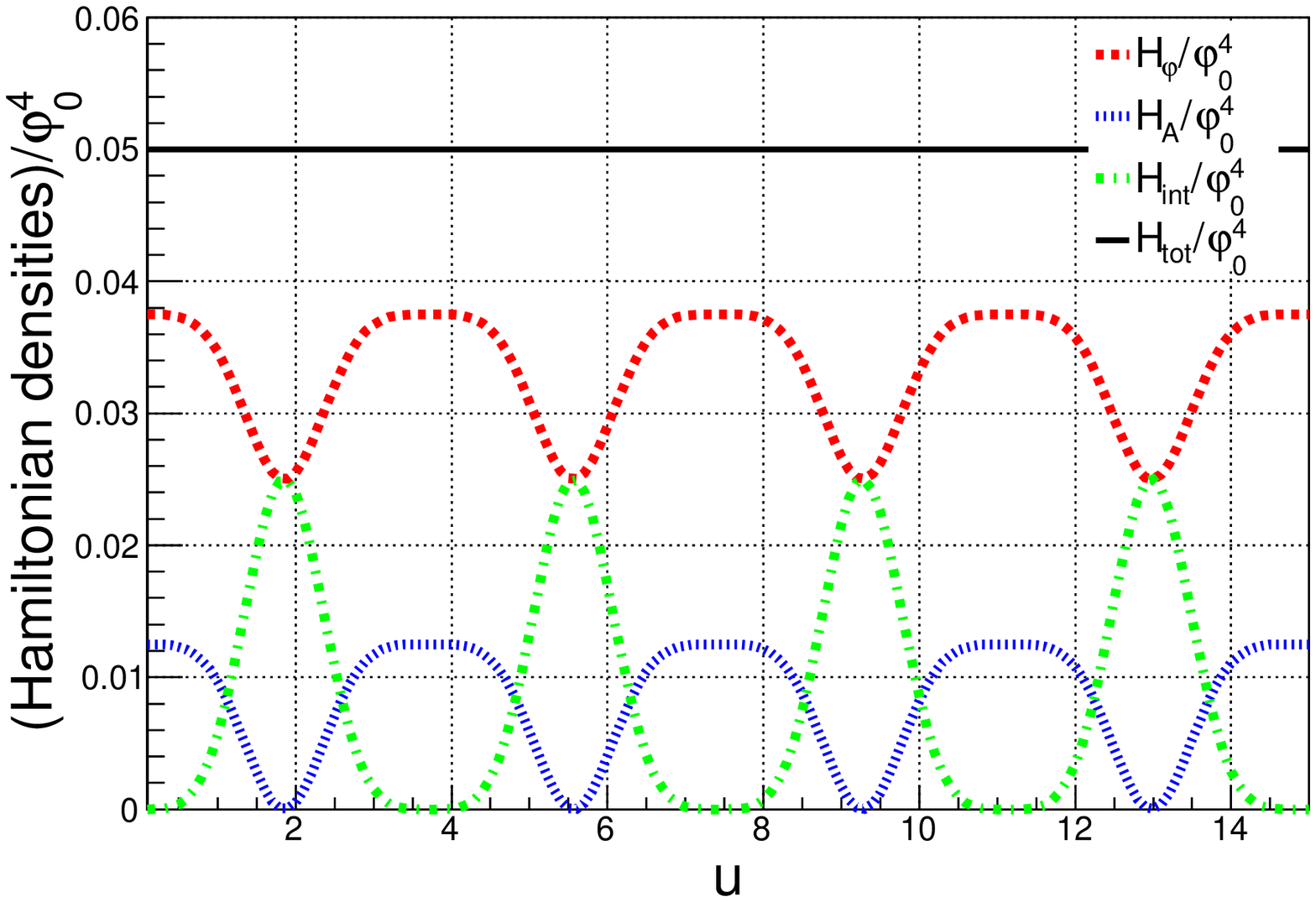}\\
   \centering
    \hspace{1.5cm}
     (b) 
 \end{minipage}
\end{tabular}
  \caption{The dimensionless plots of Hamiltonian densities for each contribution:  scalar $\mathcal{H}_{\varphi}$ (dash-line and red in color), gauge $\mathcal{H}_{A}$ (dash-dot-line and blue in color), interaction $\mathcal{H}_{\rm int}$ (dot-dot-line and green in color), and total $\mathcal{H}_{\rm tot}$ (solid-line and black in color), for (a) $cn$-type and (b) $sd$-type solutions, respectively. \label{fig.Hami.Cn.Sn}}
\end{figure}

We show each contribution of the dimensionless Hamiltonian-density for $cn$-type and $sd$-type solutions in Fig.~\ref{fig.Hami.Cn.Sn}. Each of them is calculated in the frame, $p^{\mu}=(m,\vec{0})$, for simplicity. The parameters, $\lambda=0.1$ and $e=\sqrt{0.15}$, are chosen so that we can clearly separate each contribution in the total Hamiltonian density.
  We can manifestly observe the exchange of energy density among $\mathcal{H}_{\varphi}$, $\mathcal{H}_{A}$, and $\mathcal{H}_{\rm int}$, namely, the elliptic oscillation among each contribution in the Hamiltonian densities, with keeping the constant nature of the total Hamiltonian-density in the system, as expected. Note that we fix the gauge $\chi=0$ in this paper, but we should take into account another contribution $\mathcal{H}_{\chi}$ to guarantee the gauge invariance of the total Hamiltonian density, if we change the gauge from $\chi=0$ to $\chi \neq 0$.

These Hamiltonian-densities show the periodic structure based on the Jacobi's elliptic-functions, $cn$ and $sd$, that is, $cn(u+T,k)=cn(u,k)$ and $sd(u+T,k)=sd(u,k)$. This elliptic period $T$ is given in
\begin{eqnarray}
 T = 4K\left(\frac{1}{\sqrt{2}}\right) = 7.416 \dots, \label{eq.elliptic.period}
\end{eqnarray}
where $K(k)$ is the complete-elliptic-integral of the first kind with the modulus $k$ \cite{Gradshteyn,Abramowitz}. Of course, this $T$ is not the time-period itself. To convert this into the time period $T_0$ in the rest frame of the momentum $p^{\mu}$, we should use the relation, $4K(1/\sqrt{2})=mT_0$, and then we obtain $T_0=4K(1/\sqrt{2})/m$ for a given $m$. 

Using the property of the elliptic functions, $cn(u+T/4,1/\sqrt{2})=-sd(u,1/\sqrt{2})$, the graphs of the Hamiltonian densities for $cn$-type solution coincide with those for $sd$-type solution. Alternatively, we can say that these two-solutions coincide to each other by choosing different integration-constants $\theta$ in $u$ for $cn$-type and $sd$-type solutions. 

%--- Fig.1: Hamiltonian density for CN and SN solution

%---
\subsection{Constraint on coupling parameters}
%---
Concerning the common factor which often appears in this system, $\sqrt{1-\lambda/e^2}$, we address a constraint on this factor. We should assume that the inside of the square root should be positive to avoid an imaginary solution for the gauge field. This assumption restricts the parameters in the model and thus we obtain the following constraint on the couplings,
\begin{eqnarray}
 \frac{\lambda}{e^2} < 1. \label{eq.e.lam.condition}
\end{eqnarray}
Note that the same condition for the non-Abelian case is discussed in Ref.~\citen{Brihaye1983}.

It is interesting to check whether the standard model of the elementary particle satisfies the constraint or not. First of all, we should remember that our model deviates from the original-Abelian-Higgs-model because of the absence of the quadratic term in the potential. Although our potential neither exactly coincides with the Higgs potential in the standard model nor with the original-Abelian-Higgs model, we try to substitute the observed values of $e$ and $\lambda$ into the parameters in our model.
 Then, the factor, $(1-\lambda/e^2)$, reduces to, $(1-\lambda/e^2) \simeq -0.42$ for $e=0.30$ and $\lambda=0.13$~\cite{PDGHiggs}. 
Hence, the reality condition in Eq.~(\ref{eq.e.lam.condition}) does not hold for the standard model.
The reality condition in (\ref{eq.e.lam.condition}) holds, if the system possesses a large electric-coupling or a small scalar-self-coupling.
Such a model safely guarantees the existence of the stable elliptic-solutions. Therefore, we simply assume the electric charge, $e$, and the self coupling, $\lambda$, satisfy the condition in Eq.~(\ref{eq.e.lam.condition}) in this paper.

%--- Sec.3 Electromagnetic property
\section{Electromagnetic property \label{Sec3}} 
In this section, we discuss some electromagnetic-property of this system. 

One of the important property of the Abelian-Higgs model is a consequence of the Abelian nature of the theory, namely, the fact that the electromagnetic fields are gauge invariant and hence these are observables, while the chromo-electric field and chromo-magnetic field in non-Abelian-gauge-theories are not observables. Hence, we discuss the method how we can extract the signal of free-field-value-parameter of this system, $\varphi_0$, through the electromagnetic property. 

The electric-field $\vec{E}$ and the magnetic-field $\vec{B}$ can be calculated by the field-strength $F^{\mu\nu}$. The results in the rest frame of the momentum, $p^{\mu}=(m,\vec{0})$, with the polarization, $\epsilon^{\mu}=(0,0,0,1)$, for the $cn$-type solution are given in
\begin{eqnarray}
 E^{z} &=& e\varphi^2_{0}\sqrt{1-\frac{\lambda}{e^2}}
 sn\left(u,\frac{1}{\sqrt{2}}\right) 
 dn\left(u,\frac{1}{\sqrt{2}}\right), 
\end{eqnarray}
where others, $E^{x}$, $E^{y}$, and the whole magnetic field $\vec{B}$, vanish for this polarization state. (Note that we conventionally chose $\epsilon^{\mu}=(0,0,0,1)$ as a particular polarization state to show the behavior of the electromagnetic fields for simplicity, but one can choose any polarization state.) The reason why the magnetic field vanishes in this case is simply the consequence of choosing the rest frame of $p^{\mu}$, namely, $B^{i}\propto \epsilon^{ijk}p^{j}\epsilon^{k}$ with $p^{i}=0$.

The electric-charge density $\rho$ and the electric-current density $\vec{j}$ can be derived by the electromagnetic field through the relevant Maxwell-equations, $\rho = \mbox{div} \vec{E}$ and $\vec{j}=\mbox{rot} \vec{B} - \partial \vec{E}/(\partial t)$. The electric charge/current densities in the rest frame of $p^{\mu}=(m,\vec{0})$ with $\epsilon^{\mu}=(0,0,0,1)$ for $cn$-type solution are given in
\begin{eqnarray}
 j^{z} = - e^2\varphi^3_{0} \sqrt{1-\frac{\lambda}{e^2}}
 cn^3\left(u,\frac{1}{\sqrt{2}}\right), 
\end{eqnarray}
where others, $j^{x}$, $j^{y}$, and the electric-charge-density $\rho$, vanish in this case. One can easily calculate these results for $sd$-type solution, then the results are given in:
\begin{eqnarray}
 E^{z} &=& -e\frac{\varphi^2_{0}}{\sqrt{2}}\sqrt{1-\frac{\lambda}{e^2}}
 \frac{cn\left(u,\frac{1}{\sqrt{2}}\right)} 
      {dn^2\left(u,\frac{1}{\sqrt{2}}\right)}, \\
%---
j^{z} &=& 
- \frac{e^2}{2\sqrt{2}}\varphi^3_{0} 
  \sqrt{1-\frac{\lambda}{e^2}}
 \frac{sn^3\left(u,\frac{1}{\sqrt{2}}\right)} 
      {dn^3\left(u,\frac{1}{\sqrt{2}}\right)}, 
\end{eqnarray}
where others, $E^{x}$, $E^{y}$, and the whole magnetic field $\vec{B}$, vanish for the polarization state $\epsilon^{\mu}=(0,0,0,1)$, because of the same reason mentioned in the calculation for $cn$-type solution. 

\begin{figure}[htb]
 \hspace{-1.0cm}
\begin{tabular}{cc}
  \begin{minipage}[b]{0.45\linewidth}
   \centering
     \includegraphics[scale=0.35]{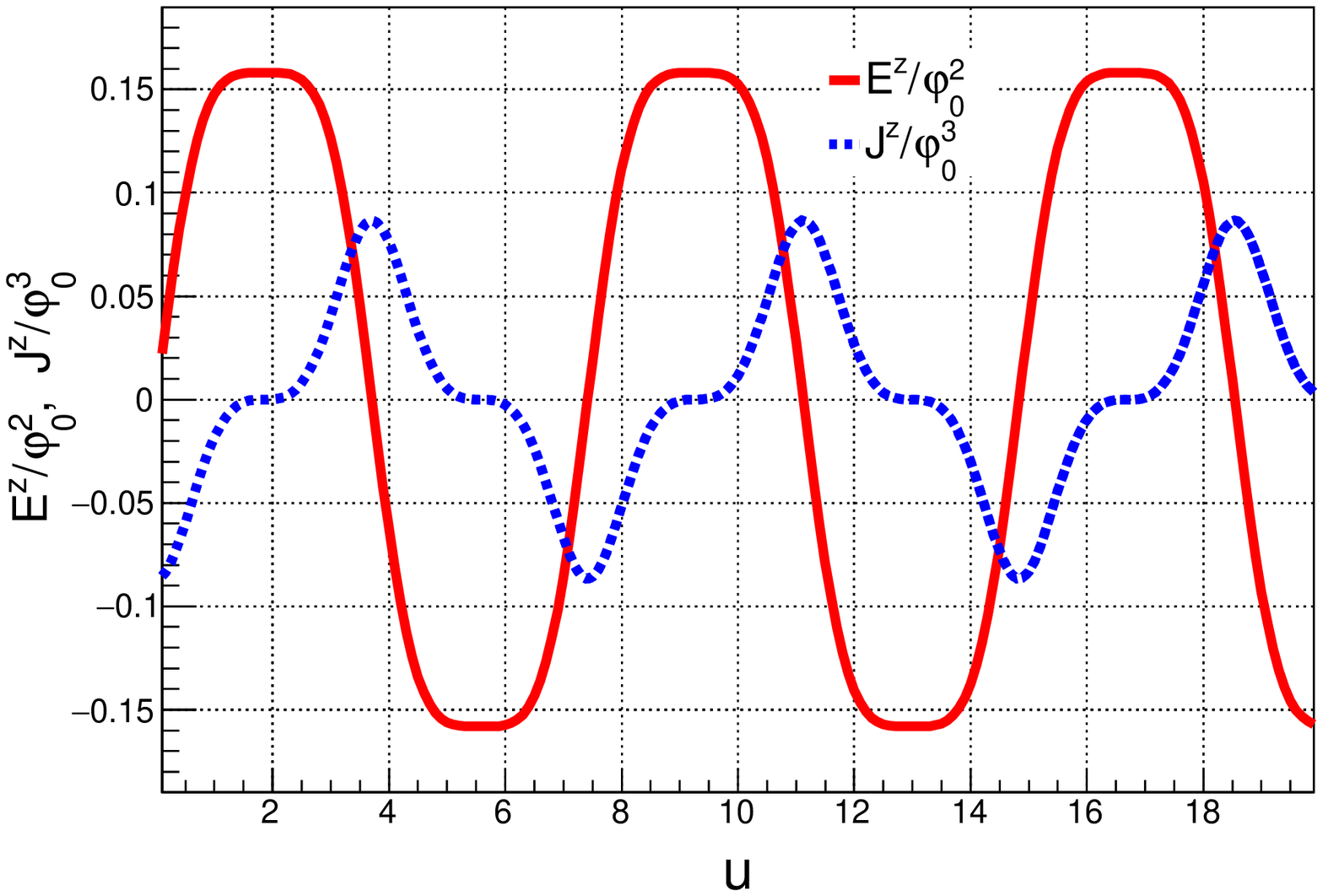}\\
    \centering
    \hspace{1.5cm}
     (a) \\
  \end{minipage}
%% SN solucion  
  \hspace{1.0cm}
  \begin{minipage}[b]{0.45\linewidth}
   \centering
     \includegraphics[scale=0.35]{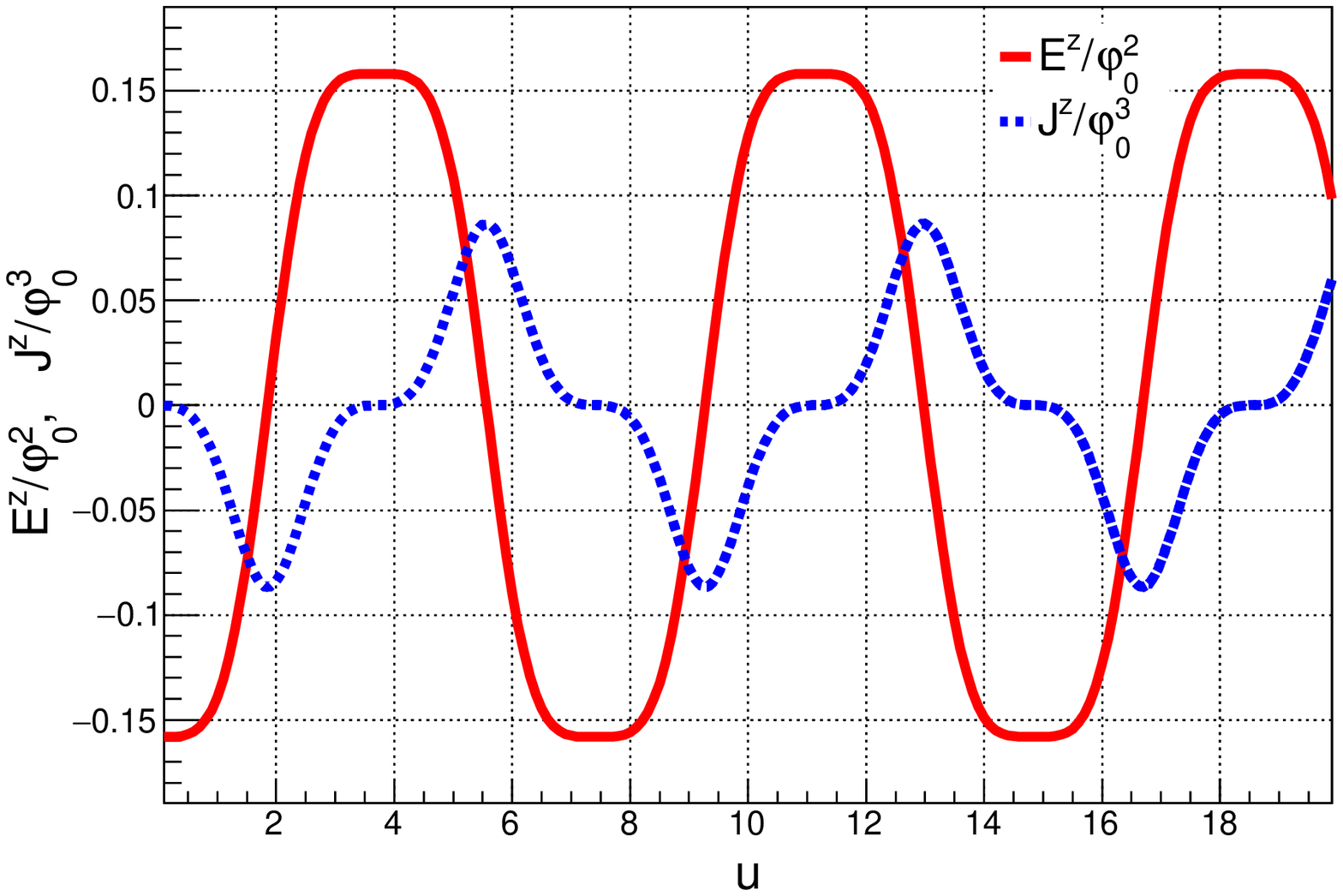} \\
    \centering
    \hspace{1.5cm}
     (b) \\
  \end{minipage}
\end{tabular}
   \caption{The dimensionless plots of the electric-field $E^{z}$ (solid-line and red in color), and of the electric-current-density $J^{z}$ (dash-line and blue in color) for (a) $cn$-type solution and (b) $sd$-type solution, respectively. \label{fig.EzJz.Cn.Sn}}
 \end{figure}

We plot the non-zero-electric-field $E^{z}$ and the electric-current-density $j^{z}$ together for $cn$-type and $sn$-type solutions in Fig.~\ref{fig.EzJz.Cn.Sn}. We here take the same parameters for $e$ and $\lambda$ used in Fig.~\ref{fig.Hami.Cn.Sn}. The reason why we plot $E^{z}$ and $j^{z}$ together in the same graph is that one can easily show the electromagnetic property expected from the Maxwell equations.
 
%--- Fig.2: Electric field and electric current density for CN and SN solution
%% CN solution
%--------------
First, one can see a periodic and sharp structure of peaks in $J^{z}/{\varphi^3_0}$, while there is periodic but less-sharp peaks like plateau in $E^{z}/{\varphi^2_0}$.  This is because that the electric-current density has more derivative than the electric field, and thus the more derivatives more increase the power of elliptic functions and hence it gives the sharp peak.

Next, one can confirm that these two-quantities show the correct behavior expected from the Maxwell equations, i.e., $j^{z} = - \partial E^{z}/(\partial t)$, in the rest frame of $p^{\mu}$. The electric-current density is zero around the region where the electric field has a periodic plateau at the maximum/minimum. This is the expected behavior from the Maxwell equations and it does not depend on either $cn$-type solution or $sd$-type solution.  In addition, the electric-current density has the maximum (minimum) in the region where the electric field crosses zero from the positive to the negative region (from the negative to the positive region). 

The periods for the electric-field and for the electric-current-density are completely fixed by the analytic nature of the Jacobi's elliptic-functions in Eq.~(\ref{eq.elliptic.period}), as we mentioned in the periodicity of the Hamiltonian density. By measuring this periodicity, one can obtain the value of $\varphi_0$, alternatively, the massive-dispersion-relation generated by the non-linear dynamics.

%--- Sec.4 Conclusion
\section{Conclusion \label{Sec4}} 
The elliptic functions as the solutions of non-linear EOMs in gauge theories have been studied intensively in the literature. In particular, one of the classical dynamics of the Higgs potential is described by the Jacobi's $dn$-type-elliptic-function. We have studied how this solution is affected by the presence of the Abelian gauge-field in this paper. As a consequence, we found the new solutions. The new solutions of the EOMs in this coupled-system is described by the Jacobi's $cn$-type-elliptic-function which has often appeared in the EOMs for pure gauge-theories. To keep the consistency of the solutions, the additional constraint on the quadratic term in the potential was derived, namely, $\mu=0$. The energy conservation is manifestly proved thanks to the oscillating property of the solutions.   

One of the interesting behavior of this system is that the generated massive-dispersion-relation is fixed by the initial-field-value $\varphi_0$ which one cannot determine theoretically. We derived the relations between this field value and the observables in this model, namely, the electromagnetic fields and the electric-charge/electric-current densities. It is found that the elliptic period of the solutions is related to the field value. Hence, we can determine this field value by measuring the period in an oscillation for these observables. 

For an implication on phenomenology, one can apply this solution to a strongly-coupled-quantum-electrodynamics, because we should take into account the phenomenological constraint on this model, namely, $\sqrt{1-\lambda/e^2}$. Another example will be found in the system expressed by a self-interacting potential, for example, inflation models. In particular, the assumption for the existence of a stable state described by the elliptic-oscillation-solution gives a constraint on the model parameters in the theories. To find out a solution with non-zero VEV is also theoretically interesting.

%--- Acknowledgments
\section*{Acknowledgments}
This work was partially funded by the Ministry of Science and Technology (MOST) of Taiwan through Grant No. MOST~111-2112-M-167-001.

%--- References ---

%--- end of main document

\begin{thebibliography}{99}
%---
\bibitem{Englert1964}
F.~Englert and R.~Brout, 
{\it Broken Symmetry and the Mass of Gauge Vector Mesons}, 
Phys.~Rev.~Lett. {\bf 13}, 321 (1964).
%---
\bibitem{Higgs1964}
P.~Higgs, 
{\it Broken Symmetries and the Masses of Gauge Bosons},
 Phys.~Rev.~Lett. {\bf 13}, 508 (1964).
%---
\bibitem{Guralnik1964}
G.~S.~Guralnik, C.~R.~Hagen and T.~W.~B.~Kibble,
{\it Global Conservation Laws and Massless Particles}, 
Phys.~Rev.~Lett. {\bf 13}, 585 (1964).
%--- 
\bibitem{Higgs.ATLAS}
G.~Aad {\it et al.} (ATLAS Collaboration),
{\it Observation of a new particle in the search for the Standard Model Higgs boson with the ATLAS detector at the LHC}, 
 Phys.~Lett.~B{\bf 716}, 1 (2012).
%--- 
\bibitem{Higgs.CMS}
S.~Chatrchyan {\it et al.} (CMS Collaboration),
{\it Observation of a new boson at a mass of $125 {\mbox{GeV}}$ with the CMS experiment at the LHC},  
Phys.~Lett.~B{\bf 716}, 30 (2012).
%--- 
\bibitem{Treat1971}
R.~P.~Treat,
{\it Plane Non-Abelian Yang-Mills Waves}, 
Nuovo Cimento {\bf 6}A, No.2, 121 (1971).
%---
\bibitem{Coleman1977}
S.~Coleman, {\it Non-Abelian plane waves},
Phys.~Lett.~{\bf 70}B, 59 (1977).
%---
\bibitem{Corrigan1977}
E.~Corrigan and D.~B.~Fairlie,
{\it Scalar field theory and exact solutions to a classical $SU(2)$ gauge theory}, 
Phys.~Lett.~{\bf 67}B, 69 (1977).
%---
\bibitem{Oh1979}
C.~H.~Oh and R.~Teh,
{\it periodic solutions of the Yang-Mills field equations}, 
Phys.~Lett.~{\bf 87}B, 83 (1979).
%---
\bibitem{Oh1985}
C.~H.~Oh and R.~Teh,
{\it Nonabelian progressive waves},
J.~Math.Phys. {\bf 26}, 841 (1985).  
%---
\bibitem{Baseyan1979}
G.~Z.~Baseyan, S.~G.~Matinyan and G.~K.~Savvidi,
{\it Nonlinear plane waves in the massless Yang-Mills theory},
Pis'ma Zh. Eksp. Teor. Fis. {\bf 29}, No.~10, 641 (1979)
%---
\bibitem{Matinyan1981}
S.~G.~Matinyan, G.~K.~Savvidi and N.~G.~Ter-Arutyunyan-Savvidi,
{\it Classical Yang-Mills mechanics. Nonlinear color oscillations},
Zh. Eksp. Teor. Fis. {\bf 80}, 830 (1981).
%---
\bibitem{Cervero1977}
J.~Cervero, L.~Jacobs and C.~R.~Nohil,
{\it Elliptic solutions of classical Yang-Mills Theory}, 
Phys.~Lett.~{\bf 69}B, 351 (1977).
%---
\bibitem{Actor1979}
A.~Actor,
{\it Elliptic function method in $\phi^4$ theory  and Yang-Mills theory},
Annals of physics, {\bf 121}, 181 (1979).
%---
\bibitem{Brihaye1983}
Y.~Brihaye,
{\it Non-Abelian plane waves in the Higgs model}, 
Lett.~Nuovo Cimento, {\bf 36}, No.9, 275 (1983). 
%---
\bibitem{Tsapalis2016}
A.~Tsapalis, E.~P.~Politis, X.~N.~Maintas and F.~K.~Diakonos,
{\it Gauss' law and nonlinear plane waves for Yang-Mills theory},
Phys.~Rev.~D{\bf 93}, 085003 (2016). 
%---
\bibitem{Achilleos2011}
V.~Achilleos et.al.,
{\it A multi-scale perturbation approach to $SU(2)$-Higgs classical dynamics: stability of nonlinear plane wave and bounds of the Higgs field mass},
Phys.~Rev.~D{\bf 85}, 027702 (2012).  
%---
\bibitem{Frasca2008}
M.~Frasca,
{\it Infrared gluon and ghost propagator},
Phys.~Lett.~B{\bf 670}, 73 (2008).
%---
\bibitem{Frasca2009}
M.~Frasca,
{\it Mapping a Massless Scalar Field Theory on a Yang-Mills Theory: Classical Case},
Mod.~Phys.~Lett.~A{\bf 24}, No.30, 2425 (2009).
%---
\bibitem{Kitadono2016} 
Y.~Kitadono and T.~Inagaki,
 {\it Elliptic oscillating solution in Higgs potential and the effects on vacuum transitions},
Phys.~Rev.~D{\bf 93}, 093014 (2016).
%--- 
\bibitem{Mohammedi2021}
N.~Mohammedi,
{\it On a classical solution to the Abelian Higgs model} 2021,
arXiv:2101.07729 [hep-th], 2021.
%--- 
\bibitem{Peskin} M.~E.~Peskin and D.~V.~Schroeder,
 {\it An Introduction to Quantum Field Theory},
 (Colorado, Westview Press, 1995).
%---
\bibitem{Itzykson} C.~Itzykson and J.-B. Zuber,
{\it Quantum Field Theory}, 
(New York, Dover Publication INC., 2005).
%--- 
\bibitem{Gradshteyn}
I.~S.~Gradshteyn and I.~M.~Ryzhik, 
{\it Table of Integrals, Series, and Products} 7th Ed.
(London, Academic Press, 2007).
%---
\bibitem{Abramowitz}
M.~Abramowitz and I.~A.~Stegun,
{\it Handbook of Mathematical Functions with Formulas, Graphs, and Mathematical Tables} 10th Ed. 
(Dover Publications, Inc., New York, 1972).
%---
\bibitem{PDGHiggs}
M.~Tanabashi (Particle Data Group),
Phys. Rev. D{\bf 98}, 030001 (2018),\\
see {\it Status of Higgs Boson Physics}.
%, \url{https://pdg.lbl.gov/2019/reviews/rpp2018-rev-higgs-boson.pdf}
\end{thebibliography}
\end{document}